%% file: main.tex
\documentclass[aps,pra,reprint,superscriptaddress,showpacs,floatfix]{revtex4-2}

\usepackage{graphicx,xcolor}
\usepackage{amsmath,amssymb,amsfonts,latexsym,mathrsfs,nccmath}
\usepackage[colorlinks=true,urlcolor=blue,citecolor=teal]{hyperref}
\usepackage{braket,siunitx}
\usepackage[caption=false]{subfig}
\usepackage{array,multirow}

\newcommand{\R}{\mathrm{R}}

\begin{document}

\title{Energy Landscape Shaping for Robust Control of Atoms in Optical Lattices}

\author{C.\ A.\ Weidner}
\email{c.weidner@bristol.ac.uk}
\affiliation{Quantum Engineering Technology Laboratories, H.\ H.\ Wills Physics Laboratory and Department of Electrical and Electronic Engineering, University of Bristol, Bristol BS8 1FD, UK}

\author{S.\ P.\ O'Neil}\email{sean.oneil@westpoint.edu}
\affiliation{Department of Electrical Engineering and Computer Science, United States Military Academy, West Point, NY 10996, USA}

\author{E.\ A.\ Jonckheere}
\email{jonckhee@usc.edu}
\affiliation{Department of Electrical and Computer Engineering, University of Southern California, Los Angeles, CA 90007, USA}

\author{F.\ C.\ Langbein}
\email{frank@langbein.org}
\affiliation{School of Computer Science and Informatics, Cardiff University, Cardiff, CF24 4AG, UK}

\author{S.\ G.\ Schirmer}
\affiliation{Faculty of Science \& Engineering, Physics, Swansea University, Swansea, SA2 8PP, UK}
\email{S.Schirmer@swansea.ac.uk, s.m.shermer@gmail.com}

\date{\today}

\begin{abstract}
Robust quantum control is crucial for realizing practical quantum technologies. Energy landscape shaping offers an alternative to conventional dynamic control, providing theoretically enhanced robustness and simplifying implementation for certain applications. This work demonstrates the feasibility of robust energy landscape control in a practical implementation with ultracold atoms. We leverage a digital mirror device (DMD) to shape optical potentials, creating complex energy landscapes. To achieve a desired objective, such as efficient quantum state transfer, we formulate a novel hybrid optimization approach that effectively handles both continuous (laser power) and discrete (DMD pixel activation) control parameters. This approach combines constrained quasi-Newton methods with surrogate models for efficient exploration of the vast parameter space. Furthermore, we introduce a framework for analyzing the robustness of the resulting control schemes against experimental uncertainties. By modeling uncertainties as structured perturbations, we systematically assess controller performance and identify robust solutions. We apply these techniques to maximize spin transfer in a chain of trapped atoms, achieving high-fidelity control while maintaining robustness. Our findings provide insights into the experimental viability of controlled spin transfer in cold atom systems. More broadly, the presented optimization and robustness analysis methods apply to a wide range of quantum control problems, offering a toolkit for designing and evaluating robust controllers in complex experimental settings.
\end{abstract}

\maketitle

\section{Introduction}\label{sec:intro}

Quantum technologies hold promise across diverse domains, including computing, cryptography, and sensing. However, realizing their full potential depends on our ability to precisely control quantum systems. Many control strategies have been proposed and excellent reviews can be found in~\cite{Rabitz2010, Glaser2015, koch_2016, koch2022, our_survey} among many others.  While dynamic control strategies involving time-varying fields have been extensively studied, they often face challenges in achieving robustness against inevitable noise and perturbations.

This work explores an alternative paradigm: \emph{energy landscape control}, where the system's Hamiltonian is carefully designed to guide its evolution towards a desired target state without time-varying controls. This offers potentially greater simplicity and intrinsic robustness, eliminating the need for complex, rapidly varying control fields, and mitigating temporal noise. (We stress that \emph{energy landscape} is different from \emph{optimization landscape}, which has been studied extensively~\cite{rabitzLandscape,mogens}). Specifically, we investigate the feasibility and robustness of energy landscape control in a realistic experimental setting: controlling spin transfer in a chain of ultracold atoms trapped in an optical lattice. This system serves as a model for quantum information processing and spin chain dynamics for quantum technologies that has been extensively studied theoretically~\cite{Bose2003,Bose2007}. Previous work on optimal control of spin chains considered time-optimal~\cite{Schirmer2015} and feedback~\cite{Schirmer2018} control, as well as more general spin network design with evolutionary algorithms~\cite{dAMico}.

These previous theoretical studies on spin chain control have often focused on idealized scenarios. In contrast, this work incorporates experimental considerations, such as realistic constraints on the optical potentials and the presence of experimental uncertainties, illustrating cases where analytical models become prohibitively complicated. We demonstrate that, despite these challenges, it is possible to design high-fidelity energy landscape controllers that achieve robust spin transfer. The required technologies have already been demonstrated, paving the way for future experimental realizations of our model. Furthermore, the methodologies developed here, including our approach to robust controller design and characterization, are broadly applicable to other quantum systems and control problems.

Simply optimizing controls for high performance (usually the fidelity of a desired protocol) is not enough: the control must maintain this performance in the presence of noise and uncertainty in system parameters, etc. One way to achieve robustness is error correction, among other error mitigation methods~\cite{RevModPhys.95.045005}. However, this involves significant overheads in the form of physical redundancy and computational costs of detecting and correcting errors. Robust quantum control can be thought of as a complement to quantum error correction that aims to design controls whose performance is inherently robust to certain types of noise or perturbations to reduce the need for error correction.

The robustness of energy landscape controllers has been considered in previous work~\cite{RIM, LCSS, XXrings} for analytical models. Here we consider a robustness analysis of these systems with an eye towards a particular experimental realization. Most experimental, computational, and theoretical implementations of robust control rely on optimization of the average fidelity or varying a single parameter and showing that the fidelity of the control remains within a suitable range. Although recent work has shown that the average fidelity is a useful practical measure of control robustness~\cite{RIM}, it requires sampling the perturbation space, which rapidly becomes computationally or experimentally expensive. Recent work has also derived methods for analytically determining the robustness of a quantum control~\cite{schirmer2024}, which are linked to sampling-based methods~\cite{LCSS}. In a similar vein, robustness measures based on second-order expansions can be used to dynamically track the effect of a given uncertainty on the control fidelity~\cite{motzoi2022}. Various analytic methods to design robust controls have also been proposed~\cite{Guerin2013, Barnes2015, sherson2018, Kestner2019, Guerin2020, Dong2023-jc, Kiely2024}. However, analytic methods are typically limited to finite-dimensional, discrete systems with well-behaved perturbations, although exceptions exist~\cite{Sugny_PMP}. Real systems, in contrast, are often described by continuous degrees of freedom (e.g., the spatial position of a laser with respect to an atom), and perturbations of experimentally relevant parameters often do not map nicely to perturbations in an analytic model. One important result of this work is to demonstrate a method for determining controller robustness that is useful for real perturbations that are not straightforward to model analytically.

Information transfer via interacting quantum spins can be implemented in a variety of platforms, including trapped ions~\cite{spintronicsIons} and electrons in quantum materials~\cite{spintronics_materials, spintronics_materials_2}. Here we focus on ultracold atoms, which are an excellent platform for many quantum technologies, from sensing~\cite{anderson2018, mitchell2022}, to qubit storage~\cite{leblanc2021}, to computing~\cite{lukin2023}. Recent work on cold atom systems also explored quantum state transfer in optical lattices~\cite{Paganelli2017} and quantum spin transistors in 1D confining potentials~\cite{Zinner2016}. For our purposes, the main advantage of cold atoms is that they can be readily cooled, trapped, and manipulated~\cite{salim2015}, and the energy landscape can be shaped directly via optical potentials. The experimental platform we consider is single atoms trapped in the individual wells of deep, periodic optical lattice potentials, where spin transport via the second-order superexchange mechanism has been studied theoretically~\cite{Lukin2003, Svistunov2003} and experimentally~\cite{Bloch2008}. In particular, we consider the optical control of spins trapped in the lattices underpinning so-called quantum gas microscopes~\cite{kuhr2010, greiner2009}, where single atoms can be trapped, imaged, and manipulated~\cite{kuhr2011, greiner2016} on the single-lattice-site level, through the use of high numerical aperture (NA) microscope objectives and spatial light modulators. Optimizing the biases between different lattice sites enables tuning of their tunneling coefficients~\cite{Bloch2008}. A method to tune these tunneling coefficients via periodic modulation of the on-site energy of the atoms in the lattice has also been recently proposed~\cite{schneider2023}. In general, quantum gas microscopes like those considered in this work are excellent analog quantum simulators, and an overview of their capabilities can be found in Ref.~\cite{bakr2021}.

This paper is structured as follows: The relevant experimental model and the theoretical underpinnings of spin transport in cold atom systems are reviewed in Sec.~\ref{sec:atom_spintronics}. Robust quantum control and the perturbations that arise in real systems are considered in Sec.~\ref{sec:perturbations}. The hybrid optimization algorithm to find the energy landscape control is presented in Sec.~\ref{sec:opt}. Sec.~\ref{sec:robust} describes the robustness evaluation of the controllers. Results are presented and discussed in Sec.~\ref{sec:results}, and Sec.~\ref{sec:conc} concludes.

\section{Quantum networking with cold atoms}\label{sec:atom_spintronics}
\subsection{Modelling spin dynamics}\label{subsec:spin_dynamics}
\subsubsection{The spin model}

In our model, we consider an ensemble of ultracold bosonic or fermionic atoms confined in a one-dimensional optical lattice potential of the form
\begin{equation}\label{eq:lattice}
    V(x) = V_\sigma \cos{(2kx + \phi)},
\end{equation}
where $k = 2\pi/\lambda$ is the wavenumber for a retro-reflected lattice of wavelength $\lambda$, and $\phi$ is the phase of the lattice. Usually, the lattice depth $V_\sigma$ is expressed in units of the recoil energy $E_{R}=\hbar^{2}k^{2}/2m$, setting $V_{\sigma} = \zeta_\sigma E_R$. Each atom is assumed to have two relevant internal states, denoted with the effective spin index $\sigma=\uparrow,\downarrow$, respectively. In general, the lattice potential can be spin-dependent~\cite{alberti2017}, depending on the wavelength of the lattice light and the internal structure of the atom, but here we consider the specific case $\zeta_\uparrow = \zeta_\downarrow$. If the atoms are at sufficiently low temperature, we can restrict our system to the lowest Bloch band and recover the Hubbard Hamiltonian
\begin{equation}\label{eq:Hamiltonian1}\begin{split}
  H_1 = &-\sum_{\langle j\ell\rangle, \sigma }\left( J_{\sigma} a_{j\sigma}^{\dag } a_{\ell\sigma }+ J_{\sigma}^\dag a_{j\sigma} a_{\ell\sigma }^\dag
\right)  \\
  & +\frac{1}{2}\sum_{j,\sigma }U_{\sigma }n_{j\sigma }\left( n_{j\sigma}-1\right) + U_{\uparrow \downarrow }\sum_{j}n_{j\uparrow }n_{j\downarrow},
\end{split}\end{equation}
where $\left\langle j,\ell\right\rangle$ denotes nearest neighbor lattice sites and $a_{j\sigma }$ are bosonic (fermionic) annihilation operators respectively for bosonic (fermionic) atoms of spin $\sigma$ localized on site $j$, and $\hat{n}_{j\sigma }=\hat{a}_{j_{\sigma }}^{\dagger }\hat{a}_{j_{\sigma }}$.

For the cubic lattice with a harmonic approximation around the minima of the potential~\cite{zoller1998}, the spin-dependent tunneling energies and on-site interaction energies are given by~\cite{Lukin2003}
\begin{align}
  J &\approx \frac{4}{\sqrt{\pi}} \, E_{R}\,  \zeta^{3/4} \exp (-2 \sqrt{\zeta}),\label{eq:J}\\
  U & \approx\frac{2\sqrt{2}}{\sqrt{\pi}} \, E_{R} \,  \zeta^{3/4} \, (k a_\mathrm{s}), \label{eq:U}
\end{align}
where in our model the scattering length $a_\mathrm{s}$ is the same for atoms of each spin $a_{\mathrm{s}, \uparrow} = a_{\mathrm{s}, \downarrow}$ and the inter-spin scattering length $a_{\mathrm{s}, \uparrow\downarrow} = a_\mathrm{s}$. Hence, $J$ and $U$ are independent of the spin state.

Following Ref.~\cite{Lukin2003}, who use a generalization of the Schrieffer-Wolff transformation~\cite{hewson} to leading order in $J_{\sigma}/U_{\uparrow \downarrow}$, it can be shown that Eq.~\eqref{eq:Hamiltonian1} is equivalent to the following effective Hamiltonian
\begin{equation}\label{eq:Hamiltonian2}\begin{split}
  H_2 =& -\sum_{\left\langle j,\ell\right\rangle }\left[ \lambda_z\sigma_{j}^{z}\sigma_{\ell}^{z}\pm \lambda_{\perp }\left( \sigma_{j}^{x}\sigma_{\ell}^{x}+\sigma_{j}^{y}\sigma_{\ell}^{y}\right) \right] \\
  &+ \sum_{j} 4\sqrt{2}\left( J_{\uparrow }^{2}/U_{\uparrow}-J_{\downarrow }^{2}/U_{\downarrow }\right) \sigma_{j}^{z},
\end{split}\end{equation}
where we define the normalized spin operators 
\begin{subequations}\begin{align}
  \hat{\sigma}_{j}^{(x)} &= \tfrac{1}{\sqrt{2}} \left( \hat{a}_{j\uparrow }^{\dagger }\hat{a}_{i\downarrow }+\hat{a}_{j\downarrow }^{\dagger}\hat{a}_{j\uparrow } \right),\\
  \hat{\sigma}_{j}^{(y)} &= -\tfrac{\imath}{\sqrt{2}} \left( \hat{a}_{j\uparrow }^{\dagger }a_{i\downarrow }-\hat{a}_{j\downarrow }^{\dagger }\hat{a}_{j\uparrow }\right),\\
  \hat{\sigma}_{j}^{(z)} &= \tfrac{1}{\sqrt{2}} (\hat{n}_{j\uparrow }-\hat{n}_{j\downarrow }).
\end{align}\end{subequations}
The $-$ ($+$) signs before $\lambda_{\perp}$ in Eq.~\eqref{eq:Hamiltonian2} correspond respectively to bosonic (fermionic) atoms. The parameters $\lambda_z$ and $\lambda_{\perp}$ are given by
\begin{equation}\label{eq:lambda}
  \lambda_{z} = \mp\frac{2J^{2}}{U}, \;\; \lambda _{\perp }=\frac{2J^2}{U},
\end{equation}
where the top (bottom) sign corresponds to bosonic (fermionic) atoms. For bosonic atoms, which we consider here, this simplifies to the following isotropic XXX Heisenberg Hamiltonian with positive effective $J$-coupling:
\begin{equation}\label{eq:Hamiltonian2_XXX_Boson}
  H_2 = \frac{2J^2}{U} \sum_{\left\langle j,\ell\right\rangle} \sigma_{j}^{z}\sigma_{\ell}^{z} + \sigma_{j}^{x}\sigma_{\ell}^{x}+\sigma_{j}^{y}\sigma_{\ell}^{y}.
\end{equation}

\subsubsection{The superexchange interaction}

To model the effect of adding energy biases, it is instructive to examine a simpler system first. Consider two atoms in a double well, where each well is occupied by a single atom and the two atoms in the system have opposite spins denoted by $\ket{\downarrow}$ and $\ket{\uparrow}$, respectively. If we add a bias term $\Delta$ to the system that corresponds to a relative ``tilt'' between the two wells, the Hamiltonian of the system is given by~\cite{Bloch2008, Lukin2003, Svistunov2003}
\begin{equation}\label{eq:BH_2well_bias}
  H = H_0 + V,
\end{equation}
where
\begin{equation}\label{eq:BH_2w_H0}
  H_0 = -\frac{\Delta}{2}\sum_{\sigma}(\hat{n}_{\sigma, L}-\hat{n}_{\sigma, R}) + U\sum_{j = L,R}\hat{n}_{j, \uparrow}\hat{n}_{j, \downarrow}
\end{equation}
and
\begin{equation}\label{eq:BH_2w_V}
  V = -J\sum_{\sigma} (\hat{a}_{\sigma, L}^\dagger \hat{a}_{\sigma, R} + \hat{a}_{\sigma, R}^\dagger \hat{a}_{\sigma, L}),
\end{equation}
where $\sigma$, as before, runs over the two spin (basis) states and $j = L, R$ corresponds to the left and right wells, respectively.

To populate the lattice with one atom per lattice site, we must operate in the Mott insulating regime. For our system, this requires $J \ll U$, or $\alpha \approx 20$~\cite{Bloch2002}. Additionally, if we assume that $J \ll |U - \Delta/2|$, we can use perturbation theory to find the second-order correction to the energies of the states in the Hilbert space defined by $\ket{n_{\uparrow, L} n_{\downarrow, L}; n_{\uparrow, R} n_{\downarrow, R}}$~\cite{Svistunov2003}. Here, we treat $H_0$ as the bare Hamiltonian and $V$ as the perturbation.

Neglecting the tunnel coupling, the ground state of $H_0$ is defined by the states $\ket{\alpha} = \ket{1~0; 0~1}$ and $\ket{\beta} = \ket{0~1; 1~0}$. The excited states we consider are the doubly-occupied states $\ket{\gamma_L} = \ket{1~1; 0~0}$ and $\ket{\gamma_R} = \ket{0~0; 1~1}$, although when considering second-order superexchange tunneling, these states act as virtual levels are remain unpopulated through the spin exchange process~\cite{Bloch2008}. The energies of the ground states are $E_\alpha = E_\beta = U$. The excited states are degenerate without the presence of $\Delta$, but the bias breaks the degeneracy, giving $E_{\gamma_L} = 2U-\Delta$ and $E_{\gamma_R} = 2U+\Delta$.

Furthermore, if we are to apply degenerate second-order perturbation theory to the states, we are interested in the matrix elements $V_{j\ell} = \braket{j|V|\ell}$. Since the perturbation is off-diagonal, $V_{jj} = 0$ and the superexchange mechanism is second-order, $V_{\alpha\beta} = 0$, only the terms that connect the ground and excited states are nonzero, and for the case considered here, $V_{ge} = V_{eg} = -J$ (since the tunnel coupling is a real parameter). Therefore, the second-order corrections to the ground state energies are given by
\begin{equation}
  J_\mathrm{eff} = \frac{2J^2U}{U^2-\Delta^2}.
\end{equation}

This can be extended to the multiple-site case by changing the biases between separate lattice sites. We use static energy landscape control to shape the effective parameters between different lattice sites to drive our desired state transfer protocols~\cite{Bloch2008}. That is, our energy landscape control modifies the biases $\Delta$ between our lattice sites, modifying our effective Hamiltonian such that the system under consideration takes the form
\begin{equation}\label{eq:Hamiltonian_effective}
  H_\mathrm{eff} =  \sum_{\left\langle j,\ell\right\rangle }J^{(j,\ell)}_\mathrm{eff}\left(\sigma_{j}^{z}\sigma _{\ell}^{z} + \sigma_{j}^{x}\sigma_{\ell}^{x}+\sigma _{j}^{y}\sigma _{\ell}^{y}\right).
\end{equation}
The effective coupling constant $J^{(j,\ell)}_\mathrm{eff}$ becomes~\cite{Lukin2003,Svistunov2003}
\begin{equation}\label{eq:J_eff}
  J^{(j,\ell)}_\mathrm{eff} = \frac{J^2 U}{U^2 - \Delta_j^2},
\end{equation}
where we define $\Delta_j$ to be the bias of site $j$ with respect to site $\ell = j + 1$, i.e., $\Delta_j = \epsilon_{j+1}-\epsilon_j$ where $\epsilon_j$ is the energy level at site $j$. $J$ and $U$ are global parameters for all sites, and they are defined by the depth of the lattice, $\zeta$. We set $\alpha = J/U$ such that Eq.~\eqref{eq:J_eff} can be written as
\begin{equation}\label{eq:J_eff_norm}
  J^{(j,\ell)}_\mathrm{eff} = \frac{2\alpha^2 U}{1-\tilde{\Delta}_j^2},
\end{equation}
where our rescaled $\tilde{\Delta}_j = \Delta_j/U$. For the simulations in this paper we consider lattice depths deep enough ($\zeta\approx20)$ so that for a 1D lattice without an appreciable external harmonic confinement, we expect to have one atom per lattice site~\cite{zoller1998}. Around these values of $\zeta$, $U\approx0.5E_R$, but it is important to note that our simulations are agnostic to the lattice depth as the $\tilde{\Delta}$ parameters are given in units of $U$. Our time is also normalized with respect to $\zeta$ so that a change in $\zeta$ simply changes the timescales and $\Delta$ values. Furthermore, note that the sign of $J^{(j,\ell)}_\mathrm{eff}$ changes from positive to negative for $\tilde{\Delta} < 1$ or $>1$, respectively; here we restrict ourselves to positive effective tunnel couplings. Finally, $J^{(j,\ell)}_\mathrm{eff}$ diverges as $\tilde{\Delta}_j\rightarrow 1$. At these values of $\tilde{\Delta}_j$, we expect the robustness of our found controllers to decrease due to the sharp slope of $J^{(j,\ell)}_\mathrm{eff}$ vs.\ $\tilde{\Delta}_j$ at these values. In what follows we relabel $\tilde{\Delta}\rightarrow\Delta$ for brevity of notation, with the understanding that the relabeled $\Delta$ refers to the scaled values.

\subsection{The problem}\label{sec:problem}

In this work, we restrict ourselves to the \emph{single-excitation subspace}, where there is only one excitation in the system. This excitation can be spread out over many atoms, but the particular problem we solve is state-to-state transfer in a spin chain with $N = 5$ sites. We choose the initial state $\ket{\psi_0} = \ket{1,0,0,0,0}$ and the desired final state $\ket{\psi_f} = \ket{0,0,0,0,1}$. We use the kets to indicate where the excitation is on the $5$-site chain, so in this case, the excitation starts at one end of the chain and is transferred to the other end of the chain after some time $T$, which is left as a free parameter to optimize with the energy landscape parameters. We neglect the effects of finite measurement time and assume that we can gain an accurate snapshot of the state $\ket{\psi(t)}$ at any time $t$. Note again that in our optimization, our transfer times are normalized such that the actual transfer time depends on the value set for the lattice depth $\zeta$.

Given this structure of the problem, we choose our performance metric to be the transfer fidelity at the transfer time $T$, defined as $F(\vec{\Delta}, T) = |\bra{\psi_f}\mathcal{U}(T) \ket{\psi_0}|^2$ where $\mathcal{U}(T) = e^{-iTH_\mathrm{eff}}$ in units with $\hbar = 1$. This expression makes explicit the dependence of the fidelity error, $\mathsf{e}(\vec{\Delta},T) = 1-F(\vec{\Delta},T)$, on $J_\mathrm{eff}^{(n)}$ and, in turn, $\vec{\Delta}$ as per Eq.~\eqref{eq:Hamiltonian_effective}, where $\vec{\Delta}$ is a vector containing the set of $\Delta_j$ for all sites $j$ considered here. In practice, we seek to minimize the fidelity error instead of maximizing the fidelity.

\subsection{The experimental model}\label{subsec:expt}

\begin{figure}
    \centering
    \includegraphics[width=\linewidth]{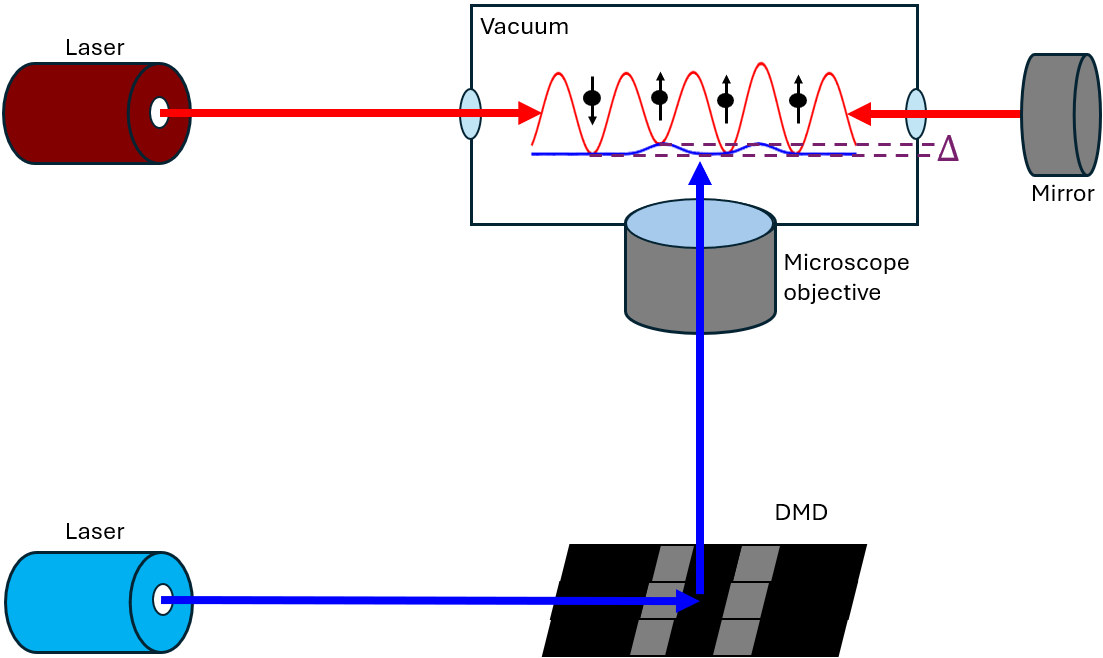}
    \caption{A cartoon diagram of the experimental setup The atoms, in a vacuum chamber, are trapped in an optical lattice potential (red) generated by a laser retro-reflecting from a mirror. In practice the trapping is 3D, but we show here a 1D lattice as this work deals only with a single dimension of the lattice potential. A second optical field (blue) generated by a laser is reflected off of a DMD and projected onto the underlying lattice potential via a high-NA microscope objective. This projected potential generates the required biases that generate the desired energy landscape for the trapped atoms, one of these biases $\Delta$ is indicated by the purple dotted lines.}
    \label{fig:experiment}
\end{figure}

The system considered here is derived from the experiment presented in Ref.~\cite{sherson2020}: a quantum gas microscopy system capable of imaging and manipulating atoms in an optical lattice potential at the single-site level. A diagram of the experimental setup is shown in Fig.~\ref{fig:experiment}.  We assume ultracold bosonic rubidium-87 atoms loaded into a cubic lattice with wavelength $\lambda_\mathrm{L} = \SI{1064}{\nano\meter}$, giving a lattice site-to-site spacing of $d = \SI{532}{\nano\meter}$. Although the experiment in Ref.~\cite{sherson2020} deals with a 3D lattice, here we model the dynamics only in 1D as we consider only spin chains.

Single-site control is achieved by placing the lattice near a microscope objective with a high NA with typical NA values between $0.6$ and $0.9$. To isolate a two-level subspace of rubidium atoms, the $\ket{F = 2, m_\mathrm{F} = -2}$ and $\ket{F = 1, m_\mathrm{F} = -1}$ hyperfine sublevels of the ground $5^2\mathrm{S}_{1/2}$ state of the atom are used, following Ref.~\cite{kuhr2011}; for this case, the lattice potential and tunneling parameters are spin independent. Furthermore, for $^{87}\mathrm{Rb}$, $a_\mathrm{s} \approx 95a_0$~\cite{hall2007, sidorov2013} for Bohr radius $a_0 = 5.29 \times 10^{-11}\text{ m}$.

Potential modification is modeled as light projected from a digital mirror device (DMD)~\cite{greiner2016, sherson2019b, gross2016} back through the microscope objective. Other spatial light modulators can also be used for such projections, e.g., liquid crystal devices~\cite{kuhr2023}. Acousto-optic deflectors can also be used for potential modification; these typically find use in the construction of optical tweezer arrays~\cite{barredo, lukin_tweezers} for, e.g., quantum computers~\cite{lukin2023} or time-averaged potentials used with bulk ultracold atomic clouds for, e.g., atomtronics~\cite{Ryu2015, atomtronics}. A review of the use of spatial light modulation devices in ultracold atom physics can be found in Ref.~\cite{yelin2021}.

For systems of this type, the radial point-spread function (PSF) at the beam focus is approximated by~\cite{gu}
\begin{equation}\label{eq:PSF}
  I_\mathrm{PSF}(\nu) = I_0 \left[\frac{2\tilde{J}_1(\nu)}{\nu}\right]^2,
\end{equation}
where $\tilde{J}_1(\nu)$ is the first-order Bessel function of the first kind with argument $\nu = 2\pi r \mathrm{NA}/\lambda$, and $r = \sqrt{x^2 + y^2}$ is the distance from the optical axis, where the beam is assumed to propagate along $z$. More generally, we define our PSF $I_\mathrm{PSF}(\nu) = |E_\mathrm{PSF}(\nu)|^2$ for some electric field $E_\mathrm{PSF}(\nu) = 2E_0 e^{i\phi(\nu)} \tilde{J}_1(\nu)/\nu$, where $E_0$ is a real scalar, and $\phi(\nu) = -kf+\nu^2/4N + \pi/2$ contains the phase information~\cite{gu}. Within the expression for $\phi(\nu)$, $k = 2\pi/\lambda$, $N$ is the Fresnel number, and $f$ is the focal length of the system.

The above PSF analysis is valid for $\mathrm{NA}<0.7$~\cite{gu}. A modification of the model, for example, including a more exact form of Eq.~\eqref{eq:PSF}, structured modifications to the PSF of the system~\cite{moerner2014, DHPSF}, or aberrations to the system, may change the exact form of the PSF but not the general robustness results presented in this work, as long as the spatial extent of the PSF is not dramatically altered. The model presented here corresponds to the intensity projected back through the potential for a single DMD pixel in the ``on'' position, in the absence of other system perturbations (including the aberrations mentioned above).

Next, we determine the general projected potential, given a DMD image represented as a binary matrix $M_\mathrm{DMD}$ with size $N_x \times N_y$ corresponding to the number of pixels on the DMD, where the zeros and ones in this matrix correspond to pixels that are ``off'' and ``on'', respectively. The intensity pattern at the atom plane is then
\begin{equation}\label{eq:conv}
  I_\mathrm{atoms}(x,y) = |E_\mathrm{PSF}(x,y) \circledast M_\mathrm{DMD}|^2,
\end{equation}
where for simplicity we have rewritten the argument $\nu$ in Eq.~\eqref{eq:PSF} in terms of Cartesian coordinates. In Eq.~\eqref{eq:conv}, $\circledast$ represents convolution and, for computational viability, we have discretized $E_\mathrm{PSF}(x,y)$ into a matrix with some spatial discretization $\delta$ in the $x$ and $y$ dimensions. In the laboratory, the field varies continuously, but the results are similar due to the small size of the DMD pixels relative to the PSF, if one takes into account the minification of the projection system. Real experimental systems typically have minification values greater than $100$ (the same as the magnification values used when imaging atoms through the other side of these microscopes)~\cite{eliasson2019}. So a $\SI{10}{\micro\meter}\times\SI{10}{\micro\meter}$ pixel is effectively $<\SI{100}{\nano\meter}$ when projected through the microscope onto the atoms, much less than a diffraction-limited spot size of a few hundred nanometers, thus in our work this approximation is valid.

Finally, the intensity scaling $I_0$ is typically found empirically. In particular, we assume that the projection light is sufficiently far detuned from resonance that losses due to atom-light scattering can be neglected, and the intensity of the projected potential is linearly proportional to the bias $\Delta_j$ felt between lattice sites. By choosing the wavelength of the projection light appropriately to the red (blue) of the dominant atomic transitions (e.g., the alkali D-lines), we can make our bias potentials attractive (repulsive). Note that the spatial extent of the PSF is also wavelength-dependent; red light leads to a PSF that has a wider spatial extent than blue light.

\section{Relevant perturbative factors}\label{sec:perturbations}

In general, given the Hamiltonian in Eq.~\eqref{eq:Hamiltonian_effective} and our model as described in Sec.~\ref{sec:atom_spintronics}, there is a set of experimental factors that can lead to perturbations in the fidelities of a given controller. We evaluate the robustness of a given controller with respect to these parameters.

The first of these drifts relates to changes in laser power, which manifests in two ways: Firstly, the projection laser power can drift. This will affect the overall magnitude of each element in $\vec{\Delta}$. For example, if the laser power drifts by a factor $\epsilon$, each element $\Delta_j$ in $\vec{\Delta}$ changes by that same amount. Secondly, a drift in the power of the lattice laser will change the lattice depth, which we defined in Sec.~\ref{subsec:spin_dynamics} as the parameter $\zeta$, which is given in units of $E_R$. This will affect the values of the bare tunneling and on-site interaction parameters $J$ and $U$ as exemplified in Eqs.~\eqref{eq:J} and~\eqref{eq:U}. However, it is important to note here that the change in the factors $J$ and $U$, as $\zeta$ changes, is not as simple as the effects of the projection laser power on the $\Delta$ values. Additionally, one must consider, when analytically calculating the sensitivity of a controller to these perturbations, that the effective tunneling potential $J_\mathrm{eff}$ as defined by Eq.~\eqref{eq:J_eff} is also non-trivial, as it is nonlinear in $J$, $U$, and $\Delta_j$. In practice, it is typically straightforward to control the laser power to at least $1$ part in $10^2$, if not more~\cite{greiner2015}.

The second set of drifts relates to defocus in the projection potential. In the model considered here, we assume that the projection beam is sent through a high-NA lens such that it focuses on a small, diffraction-limited spot at the atoms. This also means that the beam rapidly expands away from the focus, effectively leading to a wider and less intense beam at the atom potential. Thus, if the projection beam is not perfectly focused on the atoms, the $\Delta$ values change in a non-trivial way. In particular, using the same analysis as in Sec.~\ref{subsec:expt}, as the beam is defocused, the peak intensity varies as~\cite{gu}
\begin{equation}\label{eq:defocus}
  I(\zeta, \nu = 0) = I_0 \left[\frac{\sin{(\xi/4)}}{\xi/4}\right]^2,
\end{equation}
where the scaled longitudinal coordinate $\xi = \pi z \mathrm{NA}^2/2\lambda$. We further assume that the defocus is small relative to the focal length of the objective. Here, we consider defocus distances $z \approx \lambda$, so the approximations made in Eqs.~\eqref{eq:defocus} are valid. In practice, one can model defocus using Fresnel diffraction~\cite{Goodman}, which is easily done computationally. Unlike drifts in projection laser power, which changes all the $\Delta$ values in the same way, defocus affects the $\Delta$ values in a nontrivial way. Especially for projection potentials caused by complicated DMD patterns, this defocus can be highly non-trivial and difficult to describe analytically. However, defocus in an unaberrated system affects the $\Delta$ values symmetrically about the focal plane at $\xi = 0$. The sensitivity vanishes when evaluated at this point, so we do not consider this perturbation further in this work. In practice, defocus in real experimental systems tends to drift rather slowly (as it is often affected by thermal fluctuations, which occur on the sub-Hz level), and as such, they are readily controlled via occasional calibration or active feedback.

The final perturbation considered here is likely the most deleterious with regard to the experimental viability of the model considered here: the drift of the lattice potential with respect to the projection potential. That is, in typical optical lattice potentials where tunneling is considered, the spacing $d$ between adjacent lattice sites is on the order of $\SI{500}{\nano\meter}$, as larger lattice spacings cause the tunneling parameters $J$ to rapidly drop off, and $J_\mathrm{eff}$ will thus also rapidly become small. Furthermore, it is difficult to design an experiment such that the mirrors defining the position of the lattice potential are mechanically fixed with respect to those of the projection potential. Thus, as these potentials drift with respect to each other, which also typically happens on thermal time scales, the $\Delta$ values drift in a non-trivial way. As we see in Sec.~\ref{sec:results}, these perturbations substantially alter the behavior of the system. Practically, this means that if a given controller works for a well-aligned experiment, this controller is less and less effective as the lattice phase drifts. In this work, the lattice phase is set to a predetermined value with respect to the DMD, and this value is not changed during optimization, although we consider perturbations to this phase when determining controller robustness.

The above perturbations are the focus of this work. There are other perturbations that can be considered, e.g., frequency drift in the lasers, but we focus on the perturbations that are most relevant in real experiments as lasers can be precisely locked to stable cavities or atomic references, for example. Our analysis is not intended to be exhaustive but rather exemplary of the specific practical system considered here.

\section{Finding controllers}\label{sec:opt}

Finding a suitable controller requires minimizing the transfer fidelity error $\mathsf{e}(A,p,T) = |1 - F(A,p,T)|$ for all possible DMD settings $A$, laser powers $p$, and transfer times $T$, where $A$ is a matrix with binary entries defining the settings of the DMD elements. This is a hybrid control problem with both continuous control variables ($p$, $T$) subject to constraints, and a large number of binary controls --- the elements in the matrix $A$. The size of $A$ depends on the DMD array but the number of entries in $A$ can easily run into the millions, e.g., for a $1080 \times 1920$ pixel array. Due to the presence of binary control variables, this problem cannot be directly solved using gradient-based algorithms or standard quasi-Newton methods, and we found evolutionary methods to be ineffective and slow. Therefore, a two-stage process was adopted to solve the optimization problem for the controllers.

In the first step, we find optimal $\Delta_j$ values that achieve the desired process, subject to constraints on the allowed $\Delta_j$ values and the transfer time $T$. In the second step, we find the binary DMD parameters and power settings that realize these optimal $\Delta_j $ values when the potential is projected onto the underlying lattice. Our methods are described in more detail in the following subsections.

\subsection{Finding the optimal biases}\label{subsec:opt_Delta}

Solving the time-dependent Schr\"odinger equation with Hamiltonian of Eq.~\eqref{eq:Hamiltonian_effective} for a chain of length $N$ for a given set of time-invariant biases $\vec{\Delta} = \{\Delta_1,\ldots,\Delta_{N-1}\}$, an initial state $\ket{\psi_0}$, and transfer time $T$ is straightforward. Finding optimal biases $\vec{\Delta}$ that achieve transfer of a given initial state to a desired final state $\ket{\psi_f}$ at some time $T$, however, is a computationally non-trivial problem due to the complexity of the optimization landscape, especially if the transfer time and the magnitudes of the biases are heavily constrained~\cite{Schirmer2015}. In particular, finding controllers that achieve robust transfer in a sufficiently short time is challenging. 

Given the problem as defined in Sec.~\ref{sec:problem}, we cast the optimization problem as 
\begin{equation} \label{eq:optim1}
  \min_{(\vec{\Delta},T)} \left(1 - F(\vec{\Delta},T) \right)
\end{equation}
subject to three constraints. Firstly, we require $T < T_\mathrm{max}$ where $T_{\mathrm{max}}$ must be larger than any quantum speed limit to the transfer time~\cite{Deffner_2017} yet small enough to ensure sufficiently fast transfer in the face of experimental limitations like the finite lifetime of the atoms in the lattice. 

Secondly, we require the elements of $\vec{\Delta}$ to have magnitude less than $1$ to prevent large values of $J_\mathrm{eff}^{(n)}$ due to the divergence of the denominator of Eq.~\eqref{eq:J_eff_norm}; in what follows we will show that $\Delta_j$ values close to unity lead to less robust controllers. We make a slight simplification of notation here, denoting the $J^{(n+1,n)}_\mathrm{eff}$ of Eq.~\eqref{eq:J_eff_norm} by $J^{(n)}_{\mathrm{eff}}$ since we only consider the coupling between adjacent sites. 

Finally, we require the optimized control vector $\vec{\Delta}$ to be mirror symmetric about the center of the chain. Unlike the first two constraints, this constraint is not necessary based on experimental or theoretical considerations but is inspired by the symmetry of the analytic solutions for perfect state transfer presented in Ref.~\cite{christandl}. However it must be stressed that the perfect-state transfer solutions in \cite{christandl} only apply to XX spin chains, and are \emph{not} applicable to our model due to the ZZ-coupling terms in Eq.~\eqref{eq:Hamiltonian_effective}. However, imposing mirror symmetry constraints still improves the performance of the optimization algorithm, both in terms of speed of convergence and the fidelities achieved. Improved convergence speed may be due to the reduced dimensionality of the optimization space. Solutions that do not satisfy this symmetry exist but will not be considered here.

To facilitate efficient optimization, we choose a re-normalization of the time variable in units of $\tau = (\hbar \alpha_0^2)/(U \alpha^2)$ where $\alpha = J/U$ and the value of $U$ is given in multiples of the recoil energy $E_R$. The nominal values of $J$ and $U$ are taken as $0.01$ and $1$, respectively, so that $\alpha_0$ is $0.01$. This scaling permits all dependence of the solution on the lattice depth $\zeta$ to be captured by $\tau$. In this way, the effect of changing the lattice depth on the optimized controllers is accounted for by proper re-scaling of the transfer time. 

The optimization problem in Eq.~\eqref{eq:optim1} of finding the biases and transfer time needed to achieve high-fidelity state transfer subject to the given constraints, can be solved effectively using standard optimization techniques such as constrained quasi-Newton methods with careful initialization, restart, and postselection. This process can be made more efficient by imposing intelligently-chosen symmetry constraints as outlined in~\cite{Schirmer2015}.

\subsection{Finding the optimal DMD parameters}\label{subsec:DMD_opt}

Once the optimal biases have been identified, one might assume that it should easy to find the optimal DMD parameters by deconvolution. However, this is not the case due to the ill-posed nature of deconvolution, even when given the PSF as the optimal kernel. This is partially because the DMD pattern, being a binary function, assigns to each mirror element either $0$ and $1$, or on and off. The DMD pattern must then be real and discrete, but deconvolution with a complex electric field as kernel renders the results complex. Thus, a second optimization is needed to find DMD patterns that create the required energy shifts at the atom locations. For the second optimization, we choose as optimization targets the energy level shifts obtained from the quasi-Newton optimization described in Sec.~\ref{subsec:opt_Delta}. Given a vector of biases $\vec{\Delta}$ generated by a specific combination of projection light color, power, and DMD pattern, we define the optimization objective functional as the norm of the difference with the optimized potential differences of Sec.~\ref{subsec:opt_Delta}, defined as $\vec{\Delta}^*$. 

The resulting DMD optimization problem is still challenging, as the DMD search space is very large. However, we only need to control the induced potentials at the atom locations, and it turns out that these can generally be effectively controlled by a few superpixels along the $x$-axis, where we define a superpixel as a set of pixels that are controlled together, reducing the dimensionality of the DMD search space significantly. Designating $\tilde{I}$ as the integer index set of ``on'' superpixels and $p$ as the continuous variable describing the normalized power of the DMD light, we cast the optimization problem as
\begin{equation*}
  \min_{(p,\tilde{I})} \| \vec{\Delta} - \vec{\Delta}^* \|_2
\end{equation*}
subject to the constraints defined by the number and size of ``on'' superpixels and the symmetry requirement. This is a mixed optimization problem that seeks optimal values for the continuous variable $p$ and integer values in the index set $\tilde{I}$.

Genetic or pattern search algorithms can be used to solve the resulting optimization problem, but even with these simplifications, we must contend with a hybrid optimization problem as the laser power is a continuous variable. Additionally, the genetic algorithms we tested proved very inefficient with run times on the order of hours. Thus, we formulate the optimization problem as a minimization problem with integer constraints, which we are able to solve efficiently using a surrogate optimization algorithm with mixed integer constraints~\cite{surrogateopt} that yields solutions in matter of minutes.

We define a given DMD pattern in terms of the number of ``on'' superpixels, the size of the superpixel, and the wavelength of light used to generate the potential. We define the size of a single superpixel as a width in terms of the number of contiguous active pixels in the x-direction and a height in terms of the number of contiguous active pixels in the y-direction. We define the coordinate axes so that the x-axis is parallel to the long axis of the chain, the z-axis defines the light propagation axis, and the y-axis is orthogonal to these. We also enforce symmetry in that the ``on'' superpixels are required to be mirror symmetric about the midpoint of the DMD array along the x-axis. The option to enforce symmetry in the DMD pattern followed from the observation that the best fidelity values from Sec.~\ref{subsec:opt_Delta}  were generated by symmetric $\vec{\Delta}$ vectors. 

We consider two wavelengths of DMD light, a red-detuned field at $\lambda_\mathrm{red} = \SI{940}{\nano\meter}$ and a blue-detuned field at $\lambda_\mathrm{blue} = \SI{460}{\nano\meter}$. For brevity, we refer to these two options as \emph{red} and \emph{blue}, respectively. In general, the number of ``on'' superpixels is a user-defined input that takes odd integer values following from the mirror-symmetry of the DMD pattern discussed above. 

Finally though the optimal DMD pattern returned by the optimization routine might provide a local minimum for the objective $\| \vec{\Delta} - \vec{\Delta}^* \|_2 $, this does not guarantee that the $\vec{\Delta}$ induced by the DMD pattern meets the same state transfer fidelity and transfer times generated by the target $\vec{\Delta}^*$. Hence, to ensure adequate performance in a realistic read-out time we compute the time evolution of the fidelity for each $\vec{\Delta}$ produced by the DMD optimization and filter the results, keeping only those DMD-optimized controllers $\vec{\Delta}$ that achieve a user-defined fidelity threshold within a user-defined limit on the read-out time. 

\section{Evaluating controller robustness}\label{sec:robust}

As the fidelity error, $\mathsf{e}(\vec{\Delta},T) = 1-F(\vec{\Delta},T)$, is our performance metric, we seek to evaluate how well the controller $\vec{\Delta}$ maintains performance in the face of external disturbances. Said differently, we seek a metric of \emph{robustness} that captures the change in performance due to external disturbances. In keeping with earlier work on controller robustness in spintronic networks~\cite{oneil_2024_log_sens,XXrings} we employ the differential sensitivity of the fidelity error to external perturbations for this purpose. As with the optimization procedure in Sec.~\ref{sec:opt}, we decompose the sensitivity computation into two parts: the sensitivity of the fidelity error to changes in $\vec{\Delta}$ and the sensitivity of $\vec{\Delta}$ to external disturbances.

\subsection{Sensitivity to \texorpdfstring{$\vec{\Delta}$}{Δ} perturbations}

Regardless of the form of an external physical disturbance on the system, resolving the effect of changes in the control vector $\vec{\Delta}$ on $\mathsf{e}(\vec{\Delta},T)$ is fundamental. This computation of the differential sensitivity must take into account the structure associated with the Hamiltonian. As Eq.~\eqref{eq:Hamiltonian_effective} represents a Heisenberg-coupled chain, we are guaranteed that the Hamiltonian commutes with the total spin operator so that the dynamics evolve on disjoint excitation subspaces~\cite{wang_2011_spin_network_subspaces}. As we are concerned with the specific case of a single excitation in the chain, we retain only the subspace composed of the direct sum of those eigenspaces associated with the unity eigenvalue of the total spin operator. The Hamiltonian in Eq.~\eqref{eq:Hamiltonian_effective} restricted to this single excitation subspace and under the assumption of nearest-neighbor coupling then has the explicit form
\begin{equation}
  H_\mathrm{eff} = \sum_{j=1}^{N-1}J_{\text{eff}}^{(j)}S_j 
\end{equation}
with the structure matrix 
\begin{equation}\label{eq: S_n}
  S_j := \frac{1}{2} I_N + \left(E_{j,j+1} + E_{j+1,j} \right) - \left( E_{jj} + E_{j+1,j+1} \right),
\end{equation}
where $E_{j,\ell}$ is the matrix with a single one in the $(j,\ell)$ location and all other entries zero. Given the initial state $\ket{\psi_0}$ and target state $\ket{\psi_f}$, the fidelity error at time $T$ is $\mathsf{e}(\vec{\Delta},T) = 1  - \lvert \bra{\psi_f}\mathcal{U}(T)\ket{\psi_0}\rvert^2$.

Assuming the control vector $\vec{\Delta}$ differs from its nominal value $\vec{\Delta}_0$ by some small amount, we define the perturbed fidelity error as $\tilde{\mathsf{e}}(\vec{\Delta},T)$. Calculation of the differential sensitivity of this perturbed fidelity error to a change in a single $\Delta_j$ evaluated at the nominal value of the potential difference vector $\vec{\Delta}_0$ at read-out time $T$ is
\begin{multline}
  \left. \frac{\partial \tilde{\mathsf{e}}(\vec{\Delta},T)}{\partial \Delta_j} \right|_{\vec{\Delta} =  \vec{\Delta}_0} := \xi_j = \\ 
  \left. -2\Re 
   \left\{
   \bra{\psi_f} \frac{ \partial \mathcal{U}(T)}{\partial \Delta_j} \ket{\psi_0} \bra{\psi_0} \mathcal{U}^{\dagger}(T) \ket{\psi_f}  \right\} \right|_{\vec{\Delta} 
   = \vec{\Delta}_0}.
\end{multline}
Computation of $\partial \mathcal{U}(T) / \partial \Delta_j$ requires computation of the derivative of a matrix exponential at the read-out time $T$. Specifically, we have~\cite{Tsai_2003}
\begin{align}\label{eq:differential_sensitivity_long}\begin{split}
  & \left. \frac{\partial \mathcal{U}(T)}{\partial \Delta_j} \right|_{\vec{\Delta} = \vec{\Delta}_0} = \\
  & -i T \int\limits_{0}^1 e^{-iH_\mathrm{eff} T(1 - s)} \left. \left( \frac{\partial H_\mathrm{eff} }{\partial \Delta_j} \right) \right|_{\vec{\Delta} = \vec{\Delta}_0}e^{-iH_\mathrm{eff} T s}\; ds\\
  & := -i T \mathcal{K}\left( \frac{\partial H_\mathrm{eff}}{\partial \Delta_j} \right),
\end{split}\end{align}
where we introduce the integral operator $\mathcal{K}$ for brevity and suppress the explicit notation for evaluation at $\vec{\Delta}_0$ with the understanding that all derivatives are evaluated at the nominal operating point. From Eq.~\eqref{eq:J_eff} and the dependence of $J_{\text{eff}}^{(j)}$ on $\Delta_j$ (and no other $\Delta_\ell$) we have
\begin{align}\begin{split}
  \frac{\partial H_\mathrm{eff}}{\partial \Delta_j} &= \frac{\partial H}{\partial J_{\text{eff}}^{(j)}}\frac{\partial J_{\text{eff}}^{(j)}}{\partial \Delta_n} \\  &= S_j \left[ \frac{4 J^2 U \Delta_{j,0}}{(U^2 - \Delta^2_{j,0})^2} \right]   := S_j \left[ \partial \mathcal{J}_j \right], \label{eq:structure}
\end{split}\end{align}
where $\partial \mathcal{J}_j$ is the scalar derivative of $J_{\text{eff}}^{(j)}$ with respect to $\Delta_j$ evaluated at the nominal operating point. We then write the differential sensitivity of the fidelity error to a change in $\Delta_j$ more compactly as
\begin{equation}\label{eq: differential_sens_compact}\begin{split} 
  \xi_j =-2 T (\partial \mathcal{J}_j) \Im \{ \bra{\psi_f}\mathcal{K}(S_j) \ket{\psi_0} \bra{\psi_0} \mathcal{U}^\dagger(T) \ket{\psi_f} \}.
\end{split}\end{equation}

\subsection{Sensitivity to physical perturbations}

In what follows, it is important to note that the sensitivity with respect to a given parameter $A$ is most relevant in cases where the optimization of a given controller is not done with respect to that same parameter. This is due to the fact that, for a controller that has reached a local minimum in fidelity error with respect to $A$, the sensitivity $\partial \mathsf{e}(\vec{\Delta},T)/\partial A$ vanishes by definition. An analogous measure of robustness with respect to these parameters would involve a second derivative with respect to fidelity error, with larger convexities representing larger sensitivities to perturbations. Thus, the sensitivity as presented here is best computed with respect to external perturbations that are not included in the optimization. Furthermore, a nonzero sensitivity implies that, for a given perturbation, there is a direction in which the parameter can vary such that the fidelity of the controller improves. Here, we consider the magnitude of the sensitivity of the fidelity error to changes in a given parameter as a measure of controller robustness about the nominal fidelity error as done in previous work.

\begin{figure*}
  \includegraphics[width = \textwidth]{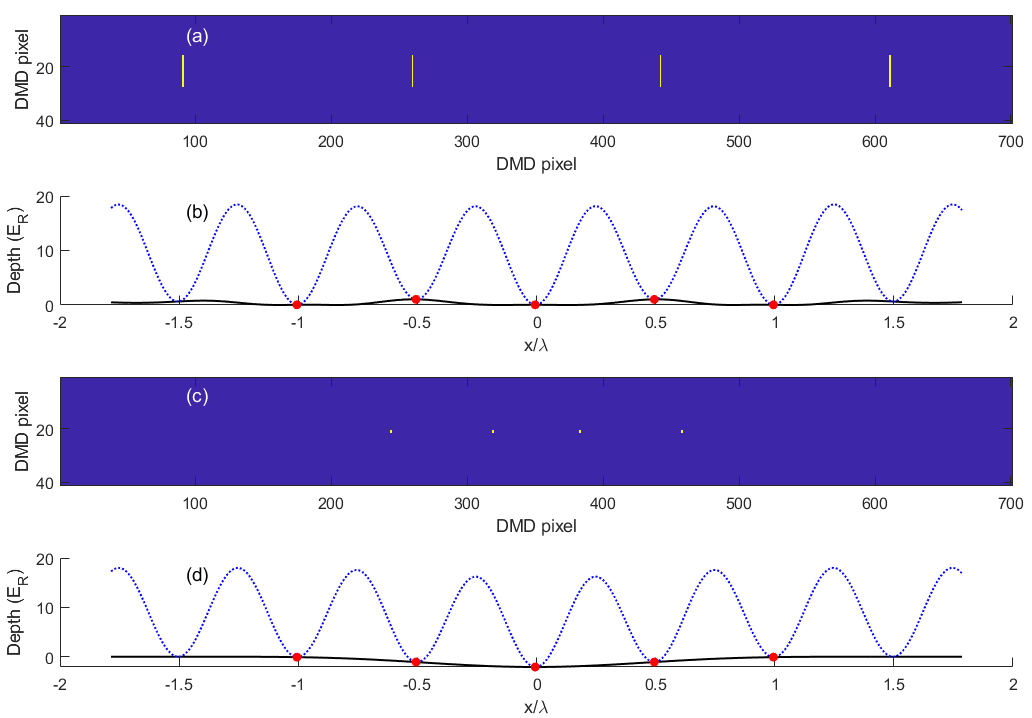}
  \caption{Example DMD patterns and the resulting potentials. Plots (a) and (c) show the DMD patterns for blue and red light where the height of the superpixel in $y$ was set to $12$ and $1$ DMD pixels, respectively (noting that the apparent difference in superpixel width in (a) is an artifact of the plotting). Plots (b) and (d) show the resulting projected potentials (black, solid line) and the total potential with a $\zeta = 18$ depth lattice on top (blue, dotted). The red dots mark the minima of the potentials used to calculate $\vec{\Delta}$. The potentials here correspond to the filled green diamond and open orange square in Figs.~\ref{fig:stuff_v_sens} and~\ref{fig:stuff_v_pow} in (a,b) and (c,d) respectively. In (a,b), the number of superpixels was taken to be $6$, and in (c,d) $4$ superpixels were used. Note that only $4$ superpixels are visible in (a) due to the plot limits, and the effect on the $\vec{\Delta}$ from the two superpixels that are not shown is minimal.
  }\label{fig:ex_DMD_potentials}
\end{figure*}

\begin{figure}
  \includegraphics[width = \columnwidth]{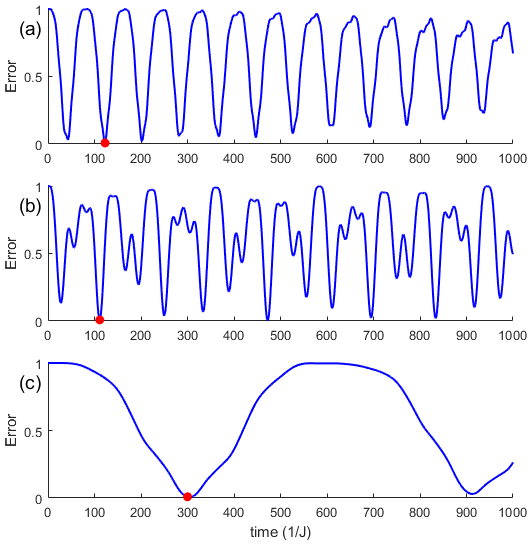}
  \caption{Fidelity error vs. time curves for three controllers that are representative of the $24$ controller dataset, showing differences in the time evolution behavior of the system. For each controller, the minimum fidelity point is shown with a red dot. The three controllers in (a)-(c) correspond to the filled blue circle, filled green diamond, and open orange square, respectively, in Figs.~\ref{fig:stuff_v_sens} and~\ref{fig:stuff_v_pow}, and they use blue-, blue-, and red-detuned potentials, respectively. The time is given in units of $1/J$; by choosing a lattice depth $\zeta$, this can be backed out to find a real, physical time. For $\zeta = 18$, a time unit of $1$ corresponds to a time of $\SI{0.186}{\milli\second}$.}\label{fig:time_traces}
\end{figure}

Given the expression for the sensitivity of the fidelity error to a change in a single $\Delta_n$, we compute the sensitivity of the fidelity error to an external disturbance $\delta$ as 
\begin{equation}
  \frac{\partial \tilde{\mathsf{e}}(\vec{\Delta},T)}{\partial \delta} = \sum_{n = 1}^{N-1} \xi_n \frac{\partial \Delta_n}{\partial \delta}.
\end{equation}
Here $\delta$ represents an infinitesimal deviation of the system from its nominal configuration. We specifically consider the two cases where $\delta$ represents drift between the lattice and the DMD array and changes in the DMD power. We characterize the first physical perturbation by the vector $\partial \vec{\Delta}/\partial x$ where the $x$-coordinate represents deviation from the nominally aligned DMD-lattice configuration along the long-axis of the chain (Deviation in the orthogonal $y$-direction yields negligibly small sensitivities relative to deviations in the $x$-direction, so they are not considered here). The other perturbation we consider is denoted by $\partial \vec{\Delta}/\partial p$ and provides the change in each $\Delta_j$ as the DMD projection intensity changes from its nominal, optimized value. We express the sensitivity to this perturbation in terms of reciprocal recoils or $1/E_R$.

\section{Results and discussion}\label{sec:results}

We consider the robustness of a set of controllers found via the two-step optimization method described in Sec.~\ref{sec:opt}. We first filter the $\vec{\Delta}$ controllers output from the first step that yield satisfactory fidelity and read-out times. We set the ceiling on the fidelity error to $10^{-2}$ and the threshold on the maximum acceptable transfer time to $\SI{130}{ms}$. This yields $16$ usable $\vec{\Delta}$ controllers that exhibit a range of sensitivity values of the fidelity error to changes in the $\Delta_j$. These $16$ controllers are then each used as the target $\vec{\Delta}^*$ for the objective functional in the DMD pixel optimization as described in Section~\ref{subsec:DMD_opt}.

For the optimization of the DMD pattern, we consider the specific experimental case described in Sec.~\ref{subsec:expt}. We optimize our DMD pattern separately for two colors of light: a red-detuned field at $\lambda_\mathrm{red} = \SI{940}{\nano\meter}$ and a blue-detuned field at $\lambda_\mathrm{blue} = \SI{460}{\nano\meter}$. For each color of light, we maintain the pixel width at a single pixel and allow the pixel height to vary from one to $25$ pixels. For each combination of superpixel size, we also optimize for the case of two, four, and six total superpixels. These choices for the superpixel size and number of superpixels are taken so that we execute a logical search upward from the smallest number of superpixels "on" and the smallest size of each superpixel to find controllers that meet the fidelity and read-out time specifications with the simplest DMD configuration. This second optimization step yields a total of $1200$ total DMD patterns for each color of light that minimizes $\| \vec{\Delta} - \vec{\Delta}^* \|$, giving $2400$ optimized controllers in total. Two example controllers and their effect on the underlying lattice are illustrated in Fig.~\ref{fig:ex_DMD_potentials}, where an example for blue and red detuning are shown. 

However, due to the complexity of the DMD optimization task, patterns that achieved $\| \vec{\Delta} - \vec{\Delta}^* \| = 0$ could not be found. Thus, the passage from direct optimization of $\vec{\Delta}^*$ to a discrete DMD pattern that approximates an optimal $\vec{\Delta}^*$ does not guarantee that the dynamics induced by $H_\mathrm{eff}$ remain unchanged. As such, we compute the time evolution of the fidelity error generated by each of the $2400$ total optimized DMD patterns and filter these for the same thresholds of $T<\SI{130}{ms}$ and $\mathsf{e} < 0.01$ used in the first step. This ultimately yields $24$ usable DMD patterns, $16$ of which are blue and eight red. Fidelity error vs. time curves for three of the $24$ controllers are found in Fig.~\ref{fig:time_traces}; these examples are representative of the entire dataset and demonstrate that the speed of dynamics can vary dramatically between controllers.

\begin{table}
\centering
\begin{tabular}{| c | c | c | c | c |}\hline
 &  \multicolumn{2}{c | }{$x$ drift} & \multicolumn{2}{c |}{$I$ drift} \\\cline{2-5}
 & $r$    & $\rho$    & $r$   & $\rho$\\\hline
$\mathrm{\min}_j|\Delta_j-1|$   &  $-0.582$ & $-0.617$   & $-0.771$  & $-0.586$\\
$T$   & $-0.382$ & $-0.400$   & $-0.206$ & $-0.336$\\
$\mathsf{e}$   & $0.150$  & $0.148$   & $0.161$  & $0.288$\\\hline
\end{tabular}
\caption{Table of Pearson $r$ and Spearman $\rho$ coefficients for the data shown in Figs.~\ref{fig:stuff_v_sens} and~\ref{fig:stuff_v_pow}, where the absolute value of the sensitivity is plotted with respect to the distance to the singularity, optimization time, and fidelity error. The only moderate to high correlations arise when the sensitivity is plotted with respect to the distance to the singularity.}\label{tab:stats}
\end{table}

\begin{figure*}
  \includegraphics[width = \textwidth]{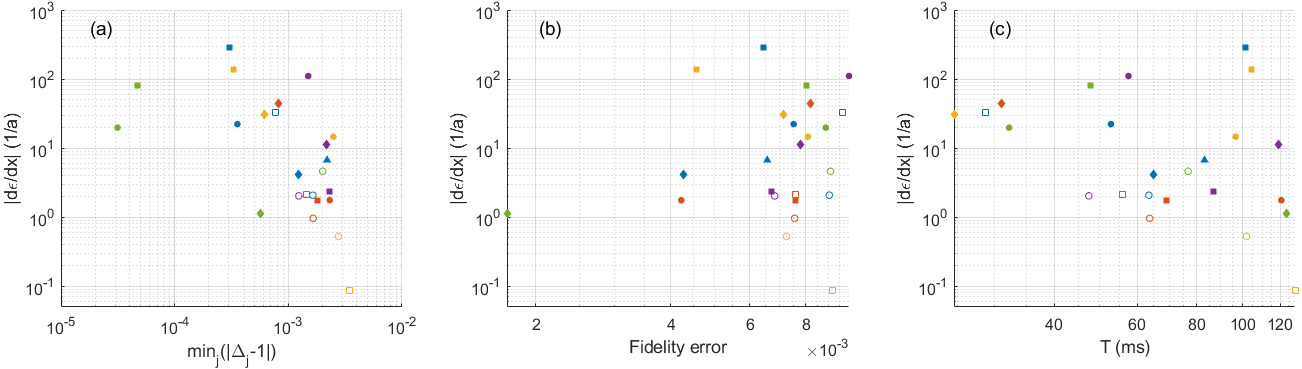}
  \caption{Results for the $24$ DMD-generated potential patterns were found for a $5$-spin chain transfer from site $1$ to site $5$. For each plot, the absolute value of the derivative of the error $\epsilon$ with respect to drifts in the projected potential along the chain direction $x$ (quantifying controller sensitivity) is shown for each controller with respect to (a) the minimum distance of a given bias $\Delta$ to the singularity at $\Delta = 1$, (b) the fidelity error, and (c) the transfer time. Filled (open) points indicate blue- (red-)detuned potentials. The distinct data points for each controller are the same between plots for ease of comparison. The units of the sensitivity are given in $1/a$, where $a$ is the lattice spacing.}\label{fig:stuff_v_sens}
\end{figure*}

\begin{figure*}
  \includegraphics[width = \textwidth]{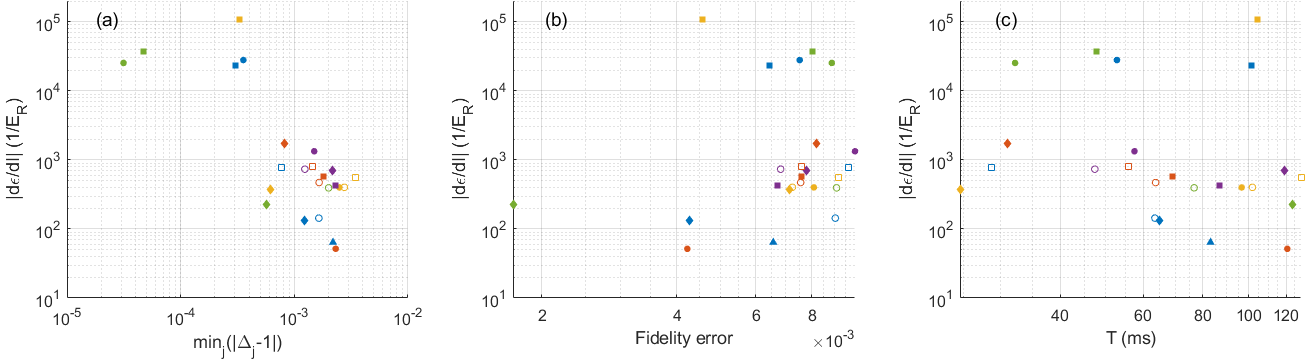}
  \caption{Same as Fig.~\ref{fig:stuff_v_sens} but showing the absolute value of the derivative of the error $\mathsf{e}$ with respect to changes in the intensity of the projected potential. The distinct data points for each controller are the same between plots and figures for ease of comparison. The units of the sensitivity are given in $1/E_\R$.}\label{fig:stuff_v_pow}
\end{figure*}

We compute the sensitivity of each of the acceptable $24$ controllers in accordance with Section~\ref{sec:robust}. The sensitivity of the fidelity error to changes in the vector $\vec{\Delta}$ is efficiently calculated analytically from Eq.~\eqref{eq: differential_sens_compact}. To compute the effect of physical disturbances given by $\partial \vec{\Delta} / \partial x$ and $\partial \vec{\Delta} / \partial p $ we opt for numerical differentiation in MATLAB. Specifically, as the DMD-optimization requires the construction of the DMD-generated potential over the entire grid of the lattice, we feed this data to MATLAB's \texttt{fit} function with \texttt{pchip} option to yield a cubic Hermite polynomial fit of the potential as a function of $x$. We then employ the \texttt{differentiate} function to evaluate the $x$ component of the potential gradient at each atom site. It then follows from linearity in the definition of $\Delta_j$, that $\partial \Delta_j / \partial x$ is given by the difference of this change in potential at atom site $j+1$ and $j$. Likewise, for the case of changes in the laser power, we build a functional dependence of $\vec{\Delta}$ on the DMD power by altering the power over the range of $0$ to $1\times E_R$. We then fit this data and compute the required derivative in the same manner as for $\partial \vec{\Delta}/\partial x$.

While these derivatives may be computed analytically or via explicit finite differences, we choose to leverage MATLAB's inherent \texttt{differentiate} function. We base this choice on the accuracy of the Piecewise Cubic Hermite Interpolating Polynomial (\texttt{pchip}) fit and efficiency. In terms of accuracy, the \texttt{pchip} fit yields near perfect goodness of fit metrics with sum-of-squares due to error less than $10^{-14}$ and R-square of unity for all fit models, so we trust the accuracy of the fit and interpolation between data points. Regarding efficiency, as the data necessary to produce the fit and evaluate the derivative is already present from the optimization, it is more time-effective to rely on the extant MATLAB code library to generate a numeric derivatives via the centered difference quotient than to either compute finite differences based on the discrete data points from scratch or compute analytic derivatives based on the coefficients in the \texttt{pchip} fit object.

In Figs.~\ref{fig:stuff_v_sens} and~\ref{fig:stuff_v_pow}, we plot the sensitivity of our $24$ controllers with respect to lattice drift in the $x$-direction and drift in the intensity of the projected potential, respectively. These parameters were chosen as they represent the two most relevant major perturbations discussed in Sec.~\ref{sec:perturbations}, but they also illustrate how the sensitivity can be used to understand the robustness of a given controller, as well as the underlying physics of the problem. For each controller, we plot the sensitivity with respect to a given parameter as a function of the minimum distance from the singularity at $\Delta_i = 1$, i.e., $ \min\limits_j {|\vec{\Delta}_j}-1|$, as well as the sensitivity with respect to fidelity error $\mathsf{e}$ and time $T$. Pearson $r$ and Spearman $\rho$ values for each set of data are shown in Table~\ref{tab:stats}. Although these values are quite consistent, they do not convey the same relational concept between two data sets. The Pearson coefficient is a correlation that assumes that the data is Gaussian. The Spearman $\rho$ is a nonparametric rank ordering coefficient; one data set (fidelity, cf. Fig.~\ref{fig:stuff_v_sens}(b)) is ordered and the Spearman quantifies how strongly the resulting ordering of the other data set (sensitivity) is concordant (+) or discordant (-) with the first one. With this in mind, we find that there is a moderate to high negative correlation between controller sensitivity and distance to the singularity, which makes physical sense in that if a controller is such that $\mathrm{min}_j|\Delta_j-1|$ is small, small changes in the parameters that govern $\vec{\Delta}$ can lead to very large changes in $\vec{\Delta}$ and, accordingly, the associated dynamics of the system. However, we find that, in general, there is very little correlation with respect to the other parameters, showing that high-fidelity solutions with high robustness to these parameters can be found. We also see that the robustness of the solution is only weakly negatively correlated with optimized transfer time. Thus, for this problem, it is most important in terms of robustness to find controllers where the $J_\mathrm{eff}$ values are reasonably stable with respect to changes in $\vec{\Delta}$.

We also find that the controllers are most sensitive to changes in the $x$-drift and power in the lattice potential, especially compared to changes in the focal plane of the projected potential. High sensitivity to changes in the $x$-drift (i.e., the lattice phase with respect to the DMD potential) is expected due to the fact that we fix the lattice phase relative to the DMD pattern at the outset, and this is not a variable parameter during optimization. However, there is a wide range of sensitivities with respect to this parameter, indicating that robust controllers can be found. Indeed, we expect that the most robust controllers are those where the maxima of the projected potentials lie at the lattice minima. If this cannot be achieved for all minima (due to diffraction limitations), the slope of the projected potential must be small at all minima, as is the case for the potentials plotted in Fig.~\ref{fig:ex_DMD_potentials}.

We do not expect high sensitivity with respect to the power in the lattice potential, as this is a parameter that we optimized for. The sensitivity we see is due to the fact that MATLAB's surrogate optimization routine cannot guarantee that it has found a local optimum with respect to the intensity. Furthermore, perturbations with respect to intensity also exhibit the highest correlation with respect to distance to the singularity, which is expected due to the direct, linear correlation between the projected intensity and $\vec{\Delta}$.

Finally, it is instructive to examine Figs.~\ref{fig:stuff_v_sens} and~\ref{fig:stuff_v_pow} to understand which controllers might be most amenable to experimental implementation. In general, one desires a controller with a low sensitivity, low fidelity error, and low transfer time; such a controller will have high performance, robustness with respect to experimental drifts, and the transfer will be completed in a reasonable time relative to typical atom lifetimes in a lattice (which can be on the order of a few seconds~\cite{Yb_QGM}). However, given the trade-off between transfer time and robustness indicated in Table~\ref{tab:stats}, we will focus on controllers with high robustness and fidelity. With this in mind, the data point indicated by the green diamonds in Figs.~\ref{fig:stuff_v_sens} and~\ref{fig:stuff_v_pow} is likely the best candidate for an experimental implementation, in no small part due to its high fidelity. However, a number of other controllers are suitable candidates, e.g., those indicated by the red circle and blue diamond.

\section{Conclusion}\label{sec:conc}

We present a method for determining the robustness of static energy landscape quantum control protocols that is suitable for use when purely analytical methods become intractable. As opposed to prior, purely theoretical work on the synthesis of energy landscape controls, we consider the physically demonstrable application of atom transport in a biased optical lattice potential based on the second-order superexchange interaction. The consideration of this system is motivated by the utility of atom-based quantum simulations. We demonstrate that the one-stage gradient-based optimization used to synthesize controllers for simple spin rings and chains is insufficient to generate high-performance controllers for this model, necessitating a two-stage mixed optimization procedure.

Specifically, we demonstrate a method to compute the sensitivity of the fidelity error to physically meaningful perturbations, including drift of the lattice with respect to the biasing potential and intensity modulation of the biasing potential. We show that in agreement with prior analytic work~\cite{spin_ring_robust_1,spin_ring_robust_2} it is possible to synthesize controllers that yield both good performance (high fidelity) and acceptable robustness (small sensitivity), in contrast to the classical control paradigm where there exists a trade-off between performance and robustness~\cite{statistical_control}. Of particular note, our analysis reveals that the most significant physical parameter that affects the sensitivity to lattice drift or power modulation is the distance of the bias potentials in $\vec{\Delta}$ to unity, where a singularity occurs in effective tunnel coupling. This bolsters the proposition that the robustness properties of quantum problems are highly system-specific, and in general such features of a given system should be considered when designing, optimizing, and evaluating controllers.

Future work will focus on extensions of these robustness analysis methods to other systems of interest in the field of quantum science and technology. In particular, future research should look beyond the energy landscape paradigm to that of systems controlled by time-dependent fields, where analytical results have already been demonstrated~\cite{schirmer2024}. Future work may also expand on the two-step optimization method, e.g., through the use of machine learning techniques~\cite{ML3}, e.g., generative neural networks~\cite{ML2, ML4}.

\input{main.bbl}

\end{document}

%% file: main.bbl
%